\documentclass[final,5p,times,twocolumn]{elsarticle}
\usepackage{epsfig}
\usepackage{amssymb}
\usepackage{lineno}

\begin{document}
\begin{frontmatter}

\title{Neutron-induced fission reaction studies with minor actinides using a double Frisch grid fission detector  }

\author{C. Yadav$^{1}$}
\author{D. Ben Simhon$^{1}$}
\author{M. Friedman$^{1}$}
\ead{moshe.friedman@mail.huji.ac.il}
\address{$^{1}$Racah Institute of Physics, Hebrew University of Jerusalem, Jerusalem 91904, Israel}

\begin{abstract} 

Neutron-induced fission cross-section studies are being pursued with a double Frisch grid fission detector. We are working towards conducting neutron-induced fission cross-section measurements of high priority $^{241}$Am($n$, $f$) and $^{244,245}$Cm($n$, $f$) fission reactions in the energy range of 1 – 2000 keV. These measurements of high priority ($n$, $f$) reactions cross-sections from the OECD Nuclear Energy Agency High Priority Request List (HPRL), are identified as reactions that require improved data in the energy range of 1-2000 keV for the safe and economic design of next-generation fast nuclear reactors. The $^7$Li($p$, $n$) reaction at various proton energies from 1.9 MeV to 3.6 MeV will be used to produce neutron flux to conduct neutron-induced fission cross-section studies. In the present special issue contribution paper, we are presenting the characterization study of the double Frisch grid fission detector. The detector is characterized with a $^{252}$Cf (spontaneous fission) source. The chamber consists of two cylindrical ionization sections in a back-to-back geometry sharing a common cathode. Each section consists of a cathode, guard rings, a grid, and an anode. The grids are made as a wire plane. The separation between the cathode and grid is 5.3 cm, and the grid to the anode is 1.2 cm. The source is mounted on a common cathode, and the chamber is operated at approximately 800 Torr pressure of P-10 gas mixture. The data acquisition with the digital data acquisition module CAEN N6725 is being set up and the preliminary analysis of the chamber performance suggests that the detector operation is good and stable. Further work with regard to employing trigger from the cathode and incorporating energy loss corrections are being pursued.

\end{abstract}


\end{frontmatter}
\section{Introduction}
\label{}

Neutron-induced fission reactions have been extensively studied since their discovery for their importance in basic and applied nuclear sciences. In basic nuclear physics, the fission cross sections provide important information on the properties of nuclear matter and help in improving the theoretical description of the fission process, while in nuclear technology they are at the basis of present and future reactor designs. Furthermore, there is a renewed interest in fission reactions in nuclear astrophysics, especially due to the important role of fission in $r$-process nucleosynthesis. 

Nuclear technology - The neutron-induced fission cross-section of the actinides is highly significant in the criticality and safety evaluation in nuclear reactors \cite{nf1}, transmutation of nuclear waste, and investigation of the nuclear fuel cycle. The design of new and innovative reactors, for safer and cleaner energy, also requires new
and higher accuracy fission cross section \cite{nf2}. The new Generation-IV reactors are expected to have, among other advantages, the capability to use nuclear
waste from currently operating reactors as fuel \cite{nf3}. The waste that is planned to be used contains mainly minor actinides, like plutonium, neptunium, americium, and curium isotopes. Gen-IV systems are expected to incinerate these isotopes and minimize nuclear waste. An alternative innovative nuclear system that can incinerate and transmute long-lived nuclear waste is the sub-critical accelerator-driven system (ADS), which uses high-energy accelerators to produce high-energy neutrons, which allows the operation of the reactor in sub-critical conditions. Apart from new reactor designs, the Th/U fuel cycle is considered for future nuclear power
options, due to the limited availability of natural uranium resources. All the above solutions, all in their R $\&$ D phase, can make nuclear power more attractive and efficient. The study of these future systems requires accurate knowledge of the cross-section of all relevant reactions, mainly the neutron-induced reactions of the minor actinides, at energies from thermal to 10s of MeV. However, significant discrepancies exist in the evaluated and measured cross-section data. The importance of accurate nuclear data is described in the High Priority Request List (HPRL) of the Nuclear Energy Agency (NEA) \cite{nf4}, and the required accuracies are summarized in \cite{nf1}. The HPRL identified several high-priority ($n$,$f$) reaction cross sections where improved data is required. Some of those within the energy range of 1-2000 keV, such as $^{241,242m}$Am($n$, $f$) and $^{244,245}$Cm($n$, $f$). $^{241}$Am (t$_{1/2}$ = 433 y) for example is one of the most important fissionable isotopes to be considered as a potential candidate for incineration/transmutation which represents about 1.8\% the actinide mass in spent fuel \cite{nf5}. 

Stellar nucleosynthesis - In explosive scenarios such as supernovae or neutron star mergers (NSM) the $r$-process nucleosynthesis takes place. During the $r$-process heavy nuclei, all the way up to unstable actinides, are formed. Spontaneous and neutron-induced fission of these actinides produce fission fragments that in turn act as seeds to a new cycle of rapid neutron capture reactions. This process is called "fission recycling" and it is considered highly significant in the $r$-process abundance distribution of the heavy elements. While the spontaneous $\beta$-delayed fission cannot be studied in the laboratory, neutron-induced fission is experimentally possible.

The neutron spectrum produced via $^{7}$Li($p$, $n$)$^{7}$Be reaction at a proton energy of 1.912 MeV has been extensively studied and employed reaction for neutron-induced reaction studies \cite{RK}.  We are working towards using this reaction at various proton energies from 1.9 MeV to 3.6 MeV to produce neutron flux in the energy range of a few keV to around 2 MeV. An important advantage of this approach is a substantial increase in the neutron intensity up to $\approx$10$^{9}$ neutrons$/$second for 100 $\mu$A proton beam. The intense proton beam from the superconducting linear RF accelerator SARAF (Soreq Applied Research Accelerator Facility) which is under upgrade (SARAF Phase II )  provides a beam current of a few mA and thus allows measurements with even higher neutron intensity on the order of 10$^{10}$ neutrons$/$second \cite{saraf, MP, MP2}. The four-order of magnitude increase in neutron intensity allows measurements of high-priority ($n$, $f$) reactions with more accuracy and precision.

A double Frisch grid fission detector \cite{3,4}, courtesy of K. E. Rehm (ANL) and M. Paul (HUJI) is being characterized to be used for these fission reaction studies. Frisch grids ionization chambers have been very successfully and widely used for neutron-induced ﬁssion-fragment studies during the past three decades, for their relative simplicity, versatility, and radiation hardness and for their capability in determining the energy and emission angle of fission-fragments with relatively good resolution \cite{5, 6, 7, 8,  9}. We plan to conduct our experiments at the PTB, Germany facility with a 10-µA proton beam at energies of 1.9-3.6 MeV on a thick Li target producing neutrons via the $^{7}$Li($p$, $n$)$^{7}$Be reaction. The target for the ($n$, $f$) reactions will be placed in the detector. The details of the detector and characterization study are described in the following sections ahead.

\section{The detector and experimental set-up }

The detector consists of two cylindrical ionization sections in a back-to-back geometry sharing a common cathode.  Each section consists of a cathode, guard rings, a grid, and an anode. The grids are made as a wire plane. The separation between the cathode and grid is 5.3 cm, and the grid to the anode is 1.2 cm.  The Frisch grid was produced by soldering 20 $\mu$m thick Au-plated tungsten wires to an annular stainless steel frame. The spacing between the wires is 1 mm. Fig. 1 shows a schematic illustration of the detector.  

\begin{figure}[h!]
\includegraphics[width= 8.5 cm]{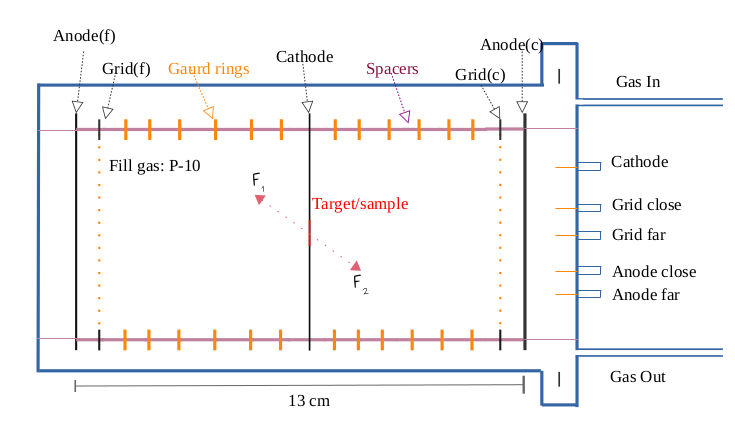}
\caption{Schematic of a double Frisch grid fission detector }
\end{figure}

\begin{figure}[h!]
\includegraphics[scale = 0.245]{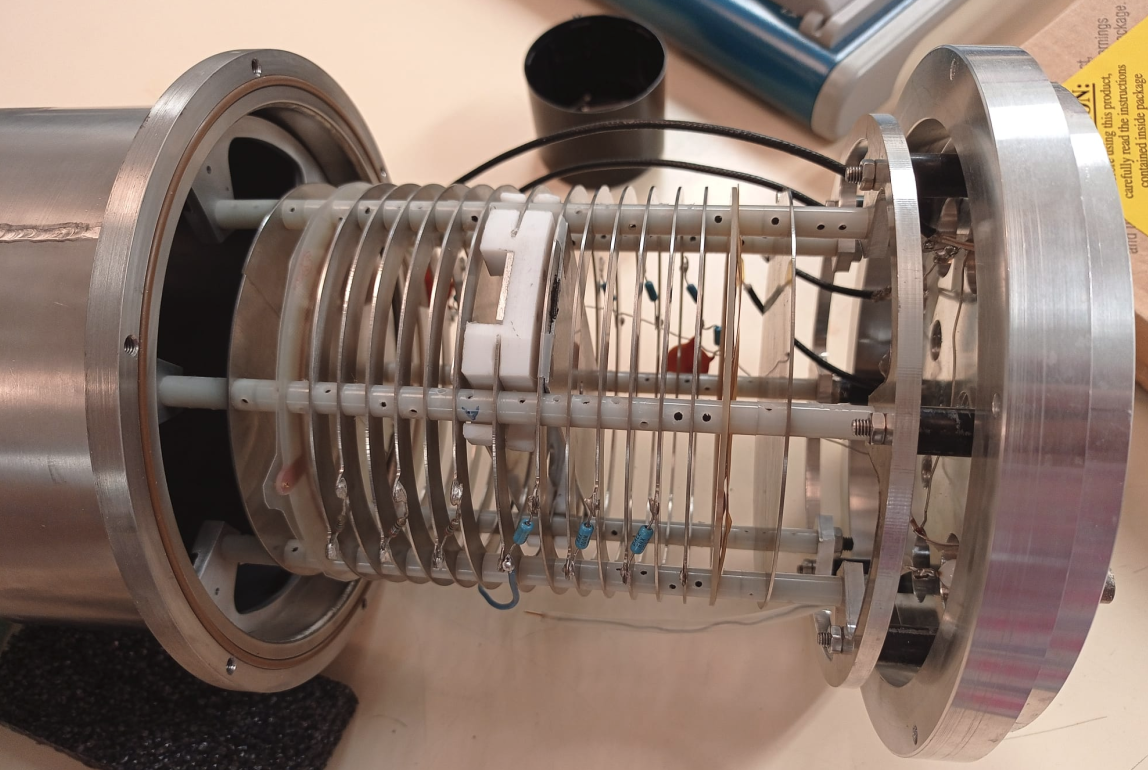}
\includegraphics[scale = 0.20]{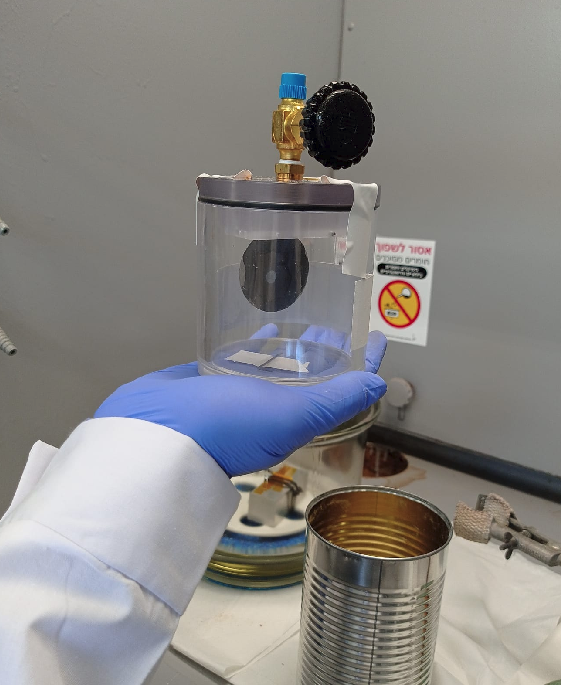}
\includegraphics[scale = 0.20]{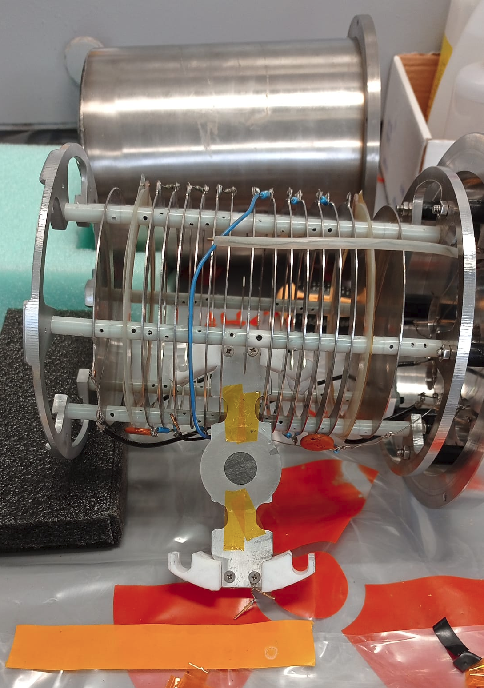}
\caption{(Top) The picture of a double Frisch grid fission detector. (Bottom left) A $^{252}$Cf (sf) source. (Bottom right)
$^{252}$Cf(sf) source being assembled on a common cathode}
\end{figure} 

\begin{figure}[h!]
\includegraphics[scale = 0.16]{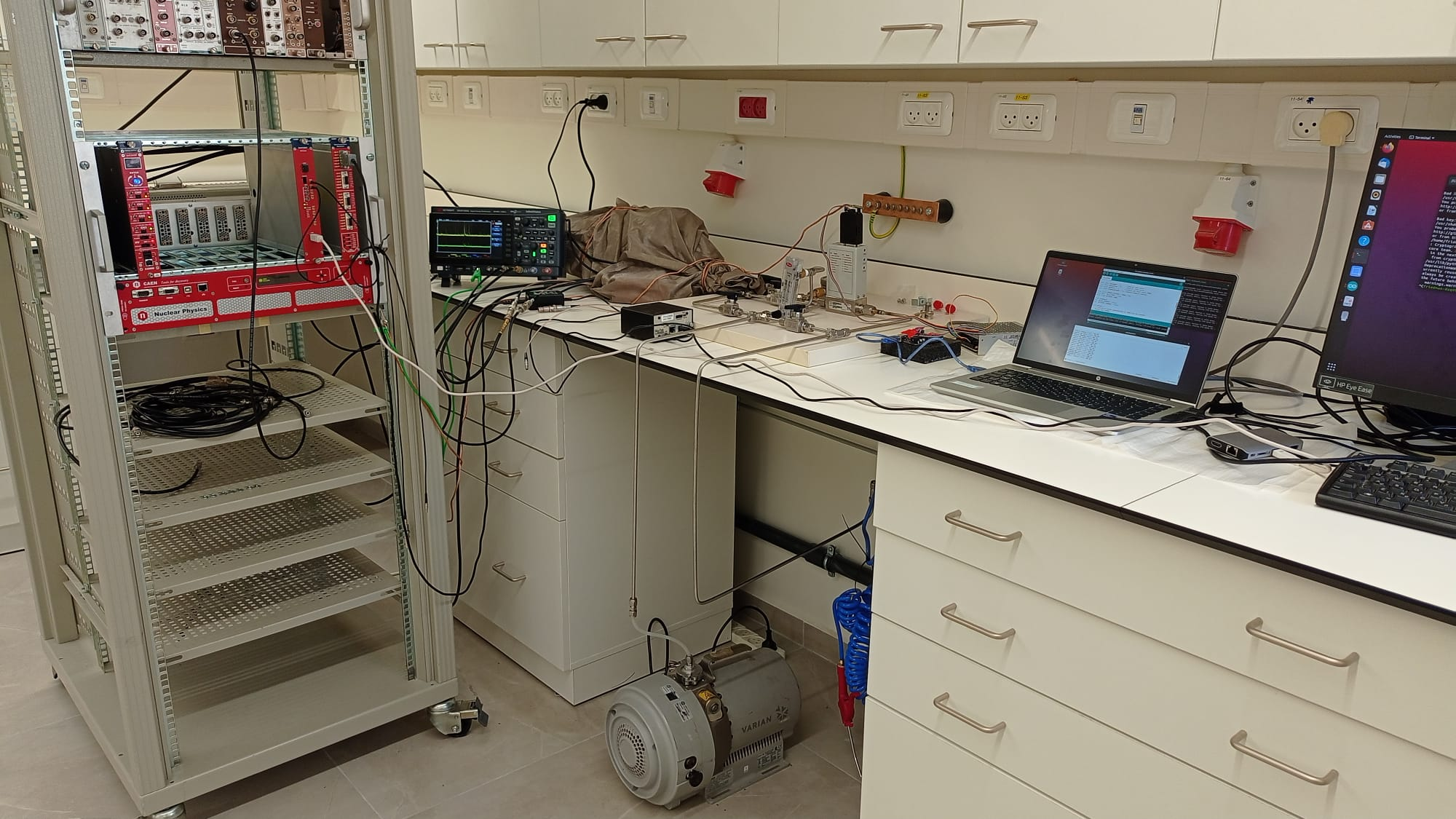}
\caption{The detector set-up.}
\end{figure}

 The double Frisch grid fission detector is characterized with a 4000 Bq  $^{252}$Cf (spontaneous fission) source deposited on a 100 $\mu$g/cm$^2$ carbon backing, a kind contribution by the Actinide Target Preparation Laboratory at Oregon State University. Fig. 2 shows the pictures of the fission detector along with a $^{252}$Cf source and the source being assembled on a common cathode.  The chamber and gas handling system were then evacuated by pumping out using an nXDS scroll pump. After reaching the suitable vacuum ($\simeq$ 0.004 Torr), gas flow to the detector was started. The chamber is operated at approximately 800 Torr pressure of P-10 gas mixture in flow mode with a flow rate of around 50 cc/min. Once the required pressure and flow rate of P-10 gas were achieved, voltages were applied gradually to the electrodes while observing the signals from the pre-amplifier through the oscilloscope.  In the present special issue contribution paper, we are reporting our preliminary results for the voltage configuration for the cathode -2200 V, the anode around +1200 V, and the grid plane was kept grounded. Furthermore, to minimize electron losses on the Frisch grid, the electric field in the grid-anode region is chosen to be of increasing strength with respect to the cathode-grid region \cite{10}. Fig. 3 shows the picture of the detector testing set-up showing the detector wrapped in copper sheets and experimental set-up.

\section{Data acquisition and preliminary results }

\begin{figure}[h!]
\includegraphics[scale = 0.2255]{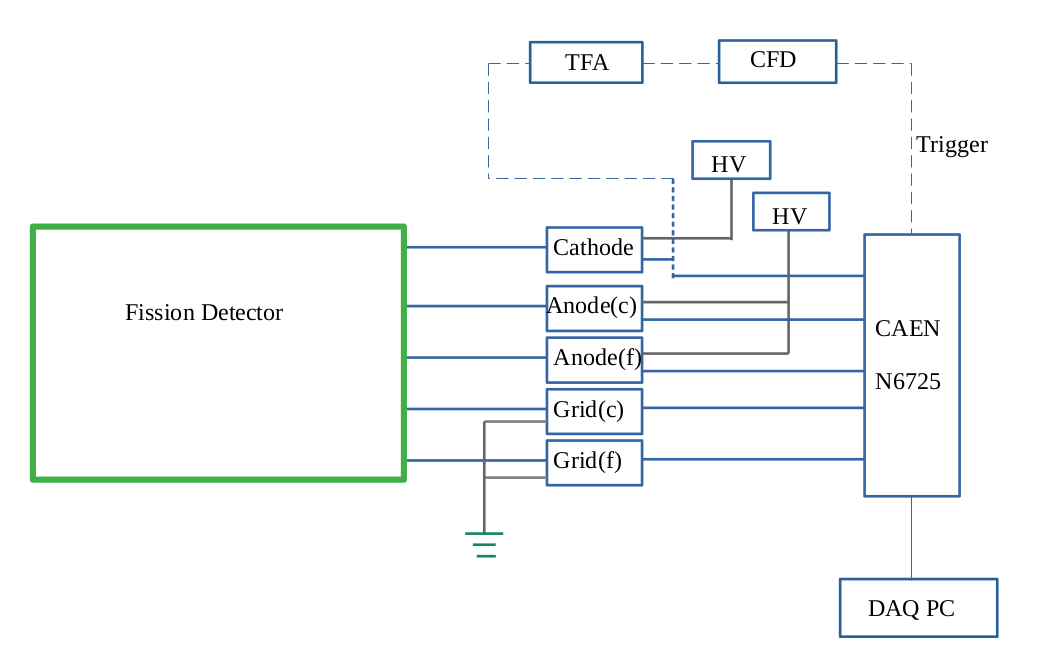}
\caption{The schematic of the digital data acquisition scheme. }
\end{figure}

\begin{figure}[h!]
\includegraphics[scale = 0.34]{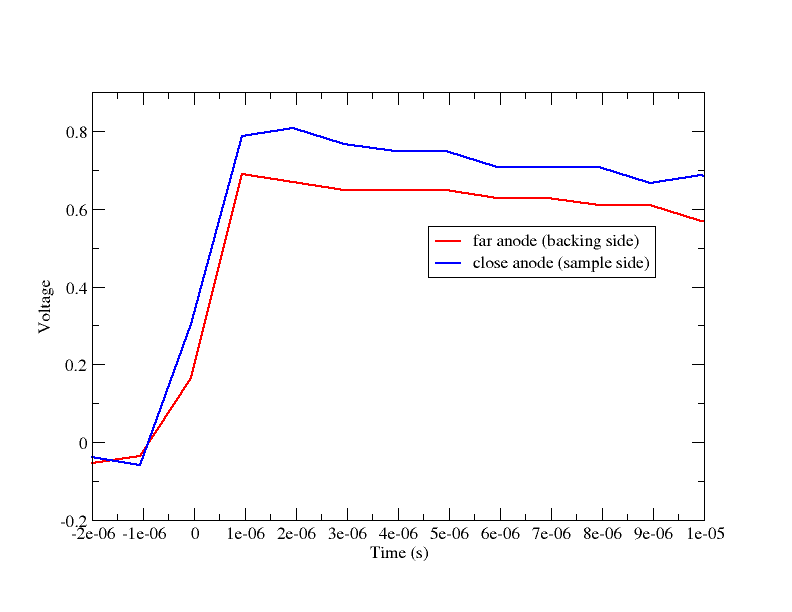}
\caption{Coincidence fission signals from the anodes. }
\end{figure}

The signals from the detector are fed to the 8-channel charge-sensitive pre-amplifier CAEN A1422C-D. The output from the pre-amplifier is further fed to the 8-channel 14-bit 250 MS/s waveform digitizer module CAEN N6725SB. Fig. 4 shows the schematic of the digital data acquisition scheme. Fig. 5 shows a typical pre-amplifier coincidence fission signal from the anodes. The  Digitizer runs a DPP-PHA (Digital Pulse Processing for Pulse Height Analysis) firmware. It uses an algorithm based on the Jordanov trapezoidal filter \cite{11}. The trapezoidal filter transforms a signal such as an exponential decay input signal from a charge-sensitive pre-amplifier into a trapezoid shape with a flat top height directly proportional to the amplitude of the input pulse. The role of the trapezoid filter is similar to that of a shaping amplifier in an analog data acquisition system. The CoMPASS \cite{12} digitizer user interface was used for communicating with the DPP-PHA firmware. The interface input can be categorized into three sections: Input Signal Settings, Trigger Settings (Timing Filter), and Energy Filter Settings. The first settings to be configured are those related to the input signal. While the trigger settings determine the criteria for a signal to be recorded and the energy filter settings control the trapezoidal fit for the incoming signals. Fig. 6 shows an example of a typical waveform from a fission event for the source backing side anode along with timing (top panel) and energy (bottom panel) filters respectively.  Fig. 7 shows pulse height distribution from the sample side anode and the backing side anode respectively at the configuration of the voltage mentioned in the previous section. Further work is being pursued with regard to connecting all the channels of the detector and employing a trigger from the cathode for data acquisition. Further work with regard to incorporating energy loss corrections for the sample and backing side spectrums is also being pursued.

\begin{figure}[h!]
\includegraphics[scale = 0.40]{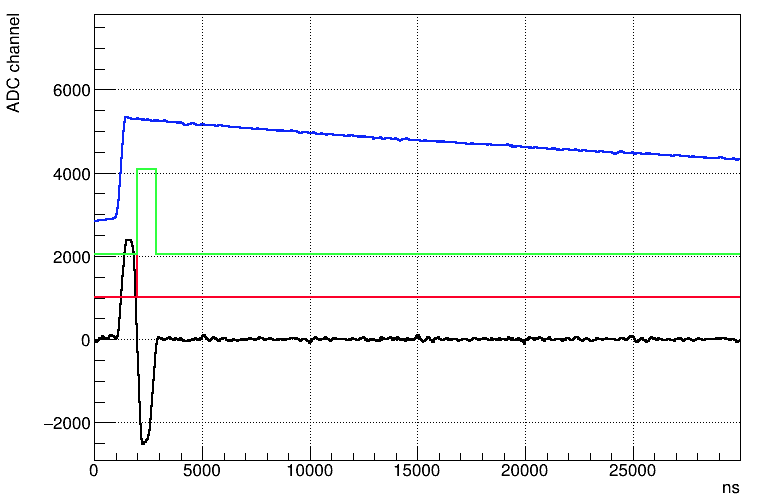}
\includegraphics[scale = 0.40]{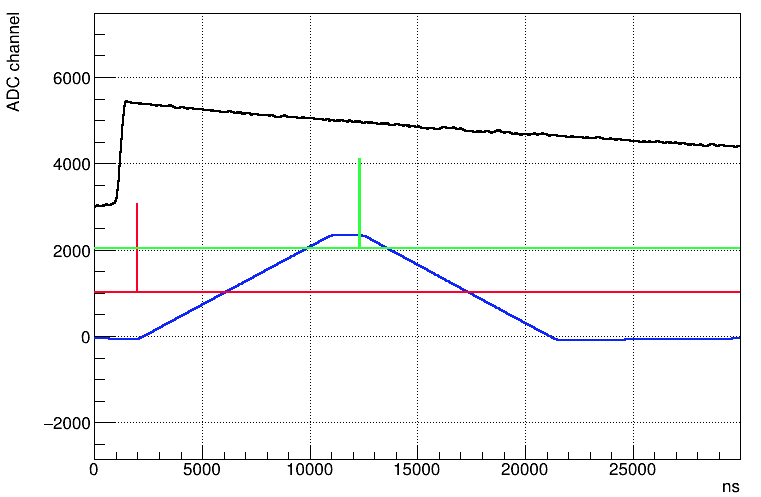}
\caption{An example of digitized far anode signal waveform with
timing (top) and energy (bottom) filters. }
\end{figure} 

\begin{figure}[h!]
\includegraphics[scale = 0.18]{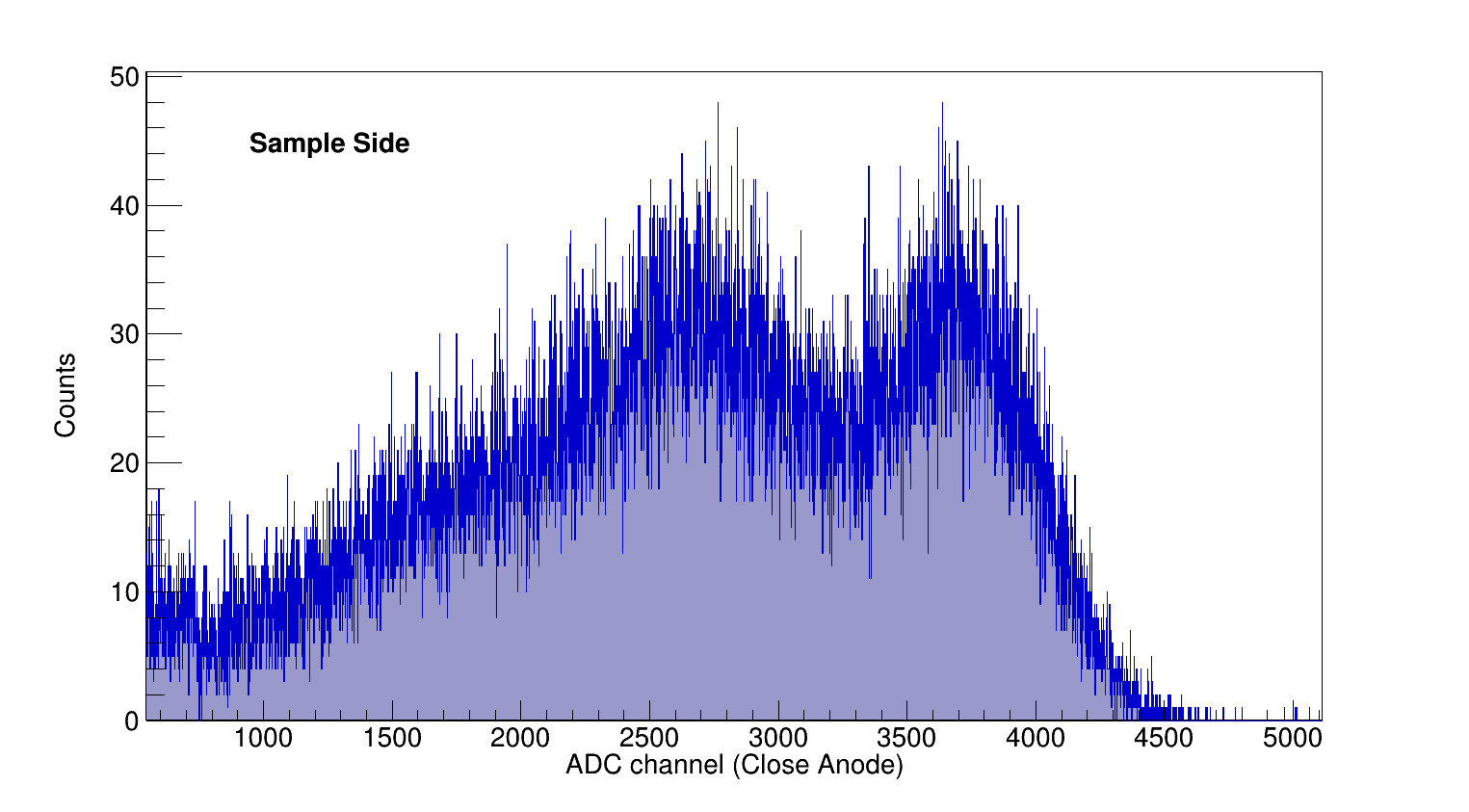}
\includegraphics[scale = 0.18]{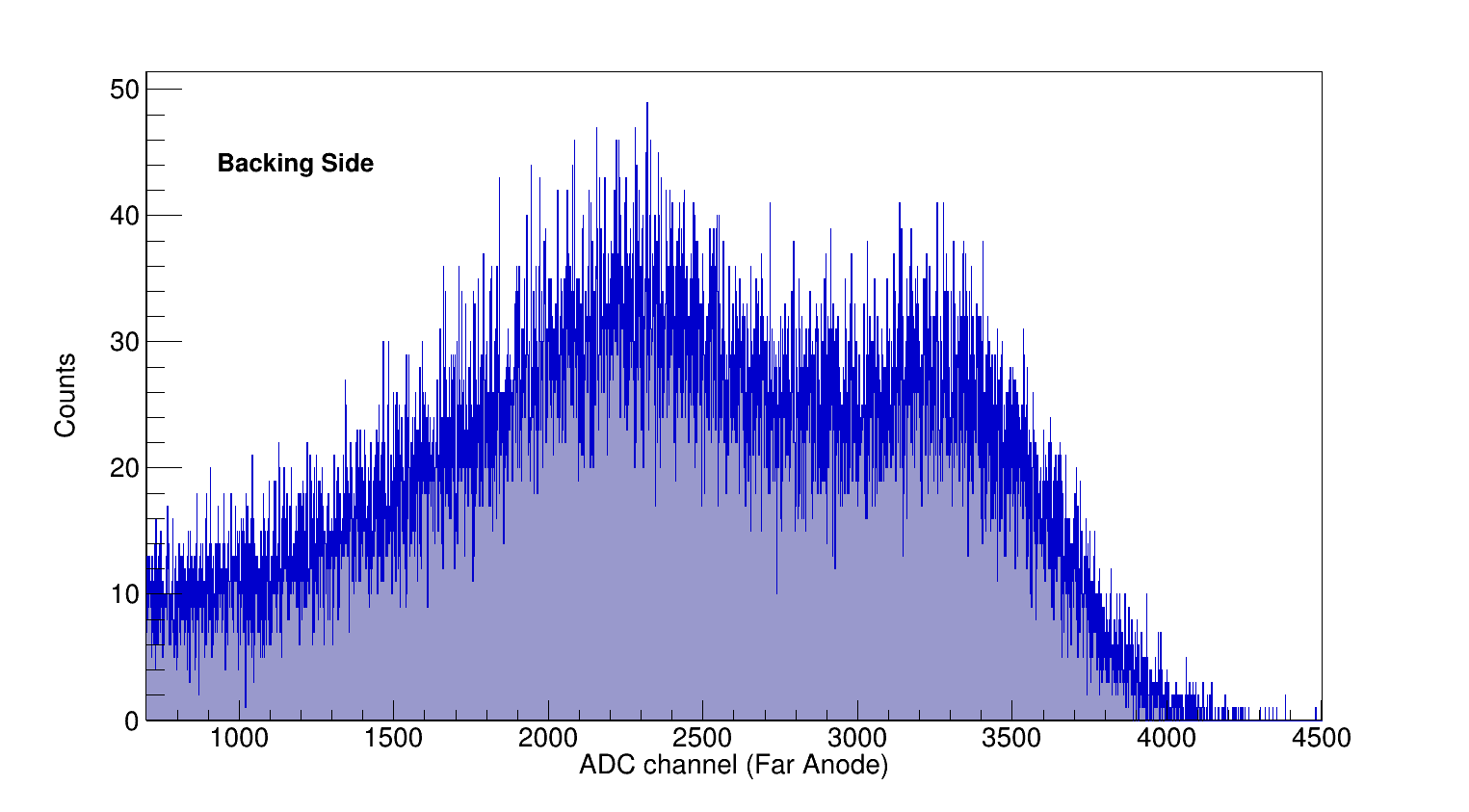}
\caption{(Top) Pulse height distribution from the sample side anode. (Bottom) Pulse height distribution from the backing side anode. }
\end{figure}

\section{Summary and conclusions}
 Neutron-induced fission cross-section studies for $^{241}$Am($n$, $f$) and $^{244,245}$Cm($n$, $f$) reactions are being pursued. In the present special issue contributory paper, our present work with regard to the testing and characterization of a double Frisch grid fission detector with a $^{252}$Cf(SF) source is discussed.  The analysis of the chamber performance suggests that the detector operation is good and stable and can be employed for neutron-induced fission reaction studies. Further work with regard to connecting all the channels of the detector and employing a trigger from the cathode and incorporating the energy loss corrections is being pursued.



\begin{thebibliography}{}



\bibitem{nf1} {M. Salvatores and R. Jacqmin, Uncertainty and Target Accuracy Assessment for Innovative Systems
Using Recent Covariance Data Evaluations, 2008.}

\bibitem{nf2}{F. Goldner and R. Versluis, Transmutation Capabilities of Generation 4 Reactors (Nuclear Energy
Agency of the OECD (NEA), 2007).}

\bibitem{nf3}{OECD Forum 2001: Sustainable Development and the New Economy (Nuclear Energy Agency of the
OECD (NEA), 2001).}

\bibitem{nf4} NEA Nuclear Data High Priority Request List, https://www.oecd-nea.org/dbdata/hprl/. 

\bibitem{nf5} {Z. Eleme \textit {et al.}, First results of the $^{241}$Am($n$,$f$) cross section measurement at the Experimental Area 2 of the n$\_$TOF facility at CERN,  EPJ Web Conf. 239, 05014 (2020).}

\bibitem{RK}{W. Ratynski, and F. Käppeler, Neutron capture cross section of $^{197}\mathrm{Au}$: A standard for stellar nucleosynthesis, Phys. Rev. C {\bf  37}, 595 (1988).}






\bibitem{saraf} {I. Mardor, O. Aviv, M. Avrigeanu  \textit {et al.}, The Soreq Applied Research Accelerator Facility (SARAF): Overview, research programs and future plans, Eur. Phys. J. {\bf A 54}, 91 (2018).}

\bibitem{MP}{ M. Paul, M. Tessler, M. Friedman \textit {et al.}, Reactions along the astrophysical s-process path and prospects for neutron radiotherapy with the Liquid-Lithium Target (LiLiT) at the Soreq Applied Research Accelerator Facility (SARAF), Eur. Phys. J. A  {\bf  55}, 44 (2019).}

\bibitem{MP2}{M. Paul, M. Tessler, M. Friedman \textit {et al.}, The liquid-lithium target at the soreq applied research accelerator facility, Eur. Phys. J. A {\bf 58}, 207 (2022).}




\bibitem{3} {X. D. Tang \textit {et al.}, New Determination of the Astrophysical S Factor S$_{E1}$ of the 12C($\alpha$, $\gamma$)$^{16}$O Reaction, Phys. Rev. Lett. {\bf 99}, 052502 (2007).}

\bibitem{4} {X. D. Tang \textit {et al.}, Determination of the $E1$ component of the low-energy $^{12}\mathrm{C}$($\ensuremath{\alpha},\ensuremath{\gamma}$)$^{16}\mathrm{O}$ cross section,  Phys. Rev. C. {\bf 81}, 045809 (2010).}

\bibitem{5} {C. Budtz-Jørgensen, \textit {et al.}, A twin ionization chamber for fission fragment detection,  Nucl. Instr. and Meth. A {\bf 258} (1987) 209.}



\bibitem{6} {M. N. Rao \textit {et al.}, Determination of fission fragment angle by the methods of grid pulse
height and the time difference between grid and collector pulses in a back-to-back gridded ionisation chamber,  Nucl. Instr. and Meth. A {\bf 313} (1992) 227-233.}

\bibitem{7} {S. Mosby \textit {et al.}, A ﬁssion fragment detector for correlated ﬁssion output studies,  Nucl. Instr. and Meth.  A {\bf 757} (2014) 75-81.}

\bibitem{8} {A. Al-Adili, \textit {et al.},  Comparison of digital and analogue data acquisition systems for
nuclear spectroscopy,  Nucl. Instr. and Meth. A {\bf 624} (2010) 684.}
\bibitem{9} {A. Al-Adili \textit {et al.}, Ambiguities in the grid-inefﬁciency correction for Frisch-Grid
Ionization Chambers,  Nucl. Instr. and Meth. A {\bf 673} (2012) 116–121.}
\bibitem{10} {O. Buneman, \textit {et al.}, Design of grid ionization chambers,  Can. J. Res.  {\bf A 27} (1949) 191.}
\bibitem{11} {V. T. Jordanov, G. F. Knoll, Digital synthesis of pulse shapes in real time for high resolution
radiation spectroscopy,  Nucl. Instr. and Meth. A {\bf 345} (1994) 337-345.}
\bibitem{12} {CoMPASS user manual (2022), https://www.caen.it/products/compass/}




\end{thebibliography}
\end{document}